\documentclass[twocolumn,amsmath,amssymb,prd]{revtex4}

\usepackage{wasysym}
\usepackage{epsfig}

\def\al{\alpha}

\def\ga{\gamma}
\def\de{\delta}
\def\ep{\varepsilon}

\def\et{\eta}

\def\la{\lambda}

\def\rh{\rho}

\def\ta{\tau}

\def\ps{\psi}
\def\om{\omega}
\def\Ga{\Gamma}

\def\Si{\Sigma}
\def\Up{\Upsilon}

\def\Om{\Omega}
\def\cA{{\cal A}}
\def\cB{{\cal B}}

\def\cF{{\cal F}}
\def\cK{{\cal K}}

\def\mn{{\mu\nu}}

\def\fr#1#2{{{#1} \over {#2}}}
\def\half{{\textstyle{1\over 2}}}

\def\frac#1#2{{\textstyle{{#1}\over {#2}}}}

\def\lsim{\mathrel{\rlap{\lower4pt\hbox{\hskip1pt$\sim$}}
    \raise1pt\hbox{$<$}}}
\def\gsim{\mathrel{\rlap{\lower4pt\hbox{\hskip1pt$\sim$}}
    \raise1pt\hbox{$>$}}}
\def\sqr#1#2{{\vcenter{\vbox{\hrule height.#2pt
         \hbox{\vrule width.#2pt height#1pt \kern#1pt
         \vrule width.#2pt}
         \hrule height.#2pt}}}}

\def\prt{\partial}

\def\etal{{\it et al.}}

\def\pt#1{\phantom{#1}}

\def\ol#1{\overline{#1}}

\def\vb#1#2{e_{#1}^{{\pt{#1}}#2}}
\def\ivb#1#2{e^{#1}_{{\pt{#1}}#2}}
\def\uvb#1#2{e^{#1#2}}

\def\ab{\overline{a}{}}

\def\cb{\overline{c}{}}

\def\sb{\overline{s}{}}

\def\twiddle{\lower4pt\hbox{\hskip-0pt{$\widetilde{}$}}}
\def\m@th{\mathsurround=0pt}
\def\cmapstochar{\mathrel{\rlap{
  \lower0.1pt\hbox{\hskip-1.75pt{$\mapstochar$}}}
  \raise0pt\hbox{\hskip2.5pt{$\twiddle$}}}}
\def\notsimfill{$\m@th\cmapstochar$}
\def\scroodle#1{\vbox{\ialign{##\crcr\notsimfill\crcr
  \noalign{\kern-4pt\nointerlineskip}
   $\hfil\displaystyle{#1}\hfil$\crcr}}}

\def\cmapstocharbig{\mathrel{\rlap{
  \lower0.1pt\hbox{\hskip0.25pt{$\mapstochar$}}}
  \raise0pt\hbox{\hskip4.5pt{$\twiddle$}}}}
\def\notsimfillbig{$\m@th\cmapstocharbig$}
\def\scroodlebig#1{\vbox{\ialign{##\crcr\notsimfillbig\crcr
  \noalign{\kern-4pt\nointerlineskip}
   $\hfil\displaystyle{#1}\hfil$\crcr}}}

\def\af{(a_{\rm{eff}})}

\def\afb{(\ab_{\rm{eff}})}

\def\afbx#1{(\ab^{#1}_{\rm{eff}})}
\def\cbx#1{\cb^{#1}}

\def\cbn{(\cb^n)}

\def\cbw{\cbx{w}}
\def\afbw{\afbx{w}}
\def\afw{(a^w_{\rm{eff}})}
\def\cw{(c^w)}

\def\lrpartial{\raise 1pt\hbox{$\stackrel\leftrightarrow\partial$}}

\def\lrDmu{\stackrel{\leftrightarrow}{D_\mu}}

\def\G{G_N}

\newcommand{\beq}{\begin{equation}}
\newcommand{\eeq}{\end{equation}}
\newcommand{\bea}{\begin{eqnarray}}
\newcommand{\eea}{\end{eqnarray}}
\newcommand{\bit}{\begin{itemize}}
\newcommand{\eit}{\end{itemize}}
\newcommand{\rf}[1]{(\ref{#1})}

\def\nwr#1{n^w_{#1}} 

\def\coef{coefficients for Lorentz violation}

\begin{document}

\title{Matter-Sector Lorentz Violation in Binary Pulsars}

\author{Ross J.\ Jennings$^1$, Jay D.\ Tasson$^{1,2}$, and Shun Yang$^1$}

\affiliation{$^1$Physics and Astronomy Department,
Carleton College, Northfield, Minnesota 55057, USA\\
$^2$Physics Department, St.\ Olaf College, Northfield, Minnesota 55057, USA}

\date{October 2015}

\begin{abstract}
Violations of local Lorentz invariance in the gravitationally coupled matter sector 
have yet to be sought
in strong-gravity systems.
We present the implications of matter-sector Lorentz violation
for orbital perturbations in pulsar systems
and show that the analysis of pulsar data can provide sensitivities to these effects
that exceed the current reach of solar system and laboratory tests by
several orders of magnitude.

\end{abstract}
\everymath{\displaystyle}

\maketitle

\section{Introduction}
\label{intro}
Since the discovery of pulsars \cite{hewish}
and the observation of binary-pulsar systems \cite{ht},
pulsars have been used to test many aspects of our current best theory
of the gravitational field,
General Relativity (GR) \cite{stairs}.
When coupled to the Standard Model (SM)
of particle physics,
the combination of GR and the SM provides a remarkable description of observed phenomena,
including the physics of pulsar systems.
Despite this success,
observational and theoretical puzzles remain,
including the lack of a satisfactory quantum-consistent theory
at the Planck scale.

Testing local Lorentz invariance
is a popular approach to the search for new physics
\cite{reviews},
including possible signals of Planck-scale physics \cite{ks}.
The gravitational Standard-Model Extension (SME) \cite{ck,akgrav}
provides a comprehensive, systematic, effective field-theory-based framework 
in which to search for Lorentz violation.
The SME can be thought of as an expansion about known physics
in Lorentz-violating operators of increasing mass dimension
contracted
with coefficients for Lorentz violation
that control the amount of Lorentz violation
in the theory.
Many theoretical \cite{lvpn,smethg} and experimental \cite{shao,he15,smeexpg}
studies have been performed in the context of the minimal 
case of operators of mass dimension 3 and 4 
in the pure-gravity sector
and analysis has now been extended beyond the minimal case
theoretically \cite{smethnm} and experimentally \cite{smeexpnm}.
The theoretical development of the gravitationally coupled matter sector \cite{lvgap,akjtprl}
has also lead to considerable experimental work \cite{he15,matterexpt}.

Over 1000 constraints on SME coefficients for Lorentz violation have now been achieved
in a wide variety of observations and experiments \cite{data},
some of which have already been achieved via the analysis of pulsar systems.
The violations of angular-momentum conservation that accompany Lorentz violation
have been exploited in the analysis of pulsar rotation rates \cite{brettpuls}.
The phenomenology of minimal Lorentz violation in the gravitational field
was considered in early work on the gravitational sector \cite{lvpn}
and constraints were considered \cite{est}.
Shao has now performed a detailed analysis based on gravity-sector phenomenology
that has lead to the best existing sensitivities to minimal gravity-sector \coef\ \cite{shao}. 
Pulsar kicks associated with Lorentz-violating neutrino physics have also been considered \cite{lamb}.
Considerable work has also been done on the special case of isotropic Lorentz violation,
or preferred-frame effects \cite{iso},
and gravitational radiation in pulsar systems \cite{wnt}
has been used to search for Lorentz violation \cite{radpuls}.
The prospects for further improvements 
in many of these pulsar-related tests using the Square Kilometre Array (SKA) \cite{ska}
are impressive.

Though much work has been done on Lorentz-violating effects in the gravitational field
in the context of pulsar systems,
the interaction of matter with the gravitational field also has implications
for the physics of gravitational systems.
This fact can be used to search for Lorentz violation in the matter sector \cite{lvgap},
including \coef\ that are unobservable in nongravitational experiments \cite{akjtprl}.
In this work,
we extend the investigation of Lorentz violation
in pulsar systems
to include the effects of Lorentz violation
in matter-gravity couplings.
In doing so,
we propose the first tests of matter-sector Lorentz violation
in strong-gravity systems
and demonstrate that the analysis of pulsar systems can extend the experimental and observational
sensitivity \cite{data,lvgap,akjtprl,he15,matterexpt} to the associated \coef\ by 
several orders of magnitude.
The search for matter-sector Lorentz violation
also involves effective Weak Equivalence Principle violation,
and sensitivity to this effect can be obtained via the proposed pulsar analysis.

This paper is organized as follows.
In Sec.\ \ref{basics}
we review the relevant theory for the analysis of matter-sector Lorentz violation
in gravitational systems,
and set up the relevant coordinates and notion for our analysis.
The leading effects of Lorentz violation in matter-gravity couplings
on binary-pulsar systems
can be though of as arising via two basic mechanisms.
We consider these in turn in Secs.\ \ref{sc} and \ref{tf},
which address secular changes in the orbit of the pulsar system
and effects on the propagation of the pulses respectively.
Section \ref{tf} also offers some discussion
of the sensitivities to Lorentz violation that might be achieved
through application of the results
to existing pulsar data.
In the final section 
we summarize the key results of the paper.
Conventions,
including the labeling of post-Newtonian orders,
are those of Refs.\ \cite{lvpn,lvgap},

\section{Basics}
\label{basics}
The description of binary-pulsar systems
is in general complicated by strong-field effects in regions close to the pulsar and companion.
However,
the systems can be modeled effectively with a post-Newtonian description
provided average orbital distances are large compared with
the radius of the bodies involved.
One approach to such a description
is the construction of a modified Einstein-Infeld-Hoffman  (EIH) Lagrangian \cite{will}.
While an EIH approach to the SME is generically of interest,
a simplified approach based on a point mass approximation
for the bodies involved was used in extracting the key features
associated with Lorentz violation in the minimal pure-gravity sector of the SME \cite{lvpn},
and the current constraints on the associated coefficients 
were found with this approach \cite{shao}.
This section provides the theoretical basics
for extending this analysis to include matter-sector effects.

\subsection{SME matter sector}

The relevant theory for the analysis to follow
is a special case of the gravitationally coupled SME
introduced in Ref.\ \cite{akgrav}
and analyzed in detail for post-Newtonian systems
in Ref.\ \cite{lvgap}.
Here we briefly review the basic theory before discussing its application
to pulsar systems.
The reader is referred to Ref.\ \cite{lvgap}
for additional detail.

At the level of the field theory of a massive Dirac fermion,
the action involving spin-independent coefficient fields for Lorentz violation $a_\mu$, $c_\mn$, and $e_\mu$
takes the form
\beq
S_\ps = 
\int d^4 x (\half i e \ivb \mu a \ol \ps \Ga^a \lrDmu \ps 
- e \ol \ps M \ps),
\label{qedxps}
\eeq
where the symbols $\Ga^a$ and $M$ 
contain the following in the present limit: 
\bea
\Ga^a
& = & 
\ga^a - c_{\mu\nu} \uvb \nu a \ivb \mu b \ga^b
- e_\mu \uvb \mu a 
\label{gamdef}
\eea
and
\beq
M
=
m + a_\mu \ivb \mu a \ga^a.
\label{mdef}
\eeq
The first term of Eq.\ \rf{gamdef}
leads to the usual Lorentz-invariant kinetic term 
for the Dirac field $\ps$, and
the first term of Eq.\ \rf{mdef}
leads to a Lorentz-invariant mass $m$.
Here $\vb{\mu}{a}$ is the vierbein,
and $e$ is its determinant.
The coefficient fields are in general particle-species dependent,
differing for protons, neutrons, and electrons
in a consideration of ordinary matter.
The fields $a_\mu$ and $e_\mu$
always appear in the combination
$\af_\mu = a_\mu + m e_\mu$
in the analysis to follow.

At leading order in Lorentz violation,
the point-particle action corresponding to \rf{qedxps}
can be written
\beq
S_u = \int d\la \left(-m  
\sqrt{-(g_\mn + 2c_\mn) u^\mu u^\nu}
-(a_{\rm eff})_\mu u^\mu \right),
\label{actionsec}
\eeq
where $\la$ is a path parameter
and $u^\mu = dx^\mu/d\la$.
Proper time is defined by $d \ta = \sqrt{-g_\mn dx^\mu dx^\nu}$.
The analysis to follow will involve macroscopic bodies
such as stars that consist of many fermions.
The leading effects of matter-sector Lorentz violation
on a macroscopic body can be achieved via the replacements
\bea
\nonumber
m & \rightarrow & \sum_w N^w m^w\\
\nonumber
c_\mn & \rightarrow & \fr{\sum_w N^w m^w \cw_\mn}{\sum_w N^w m^w}\\
\af_\mu & \rightarrow & \sum_w N^w \afw_\mu.
\label{composite}
\eea
Here sums over $w$ are over particle species,
$N^w$ denotes the number of particles of type $w$ in the body,
and $m^w$ denotes the mass of particle species $w$.
With this approach,
the pulsar and companion are modeled as point objects
coupled to matter-sector Lorentz violation
based on their respective particle-species content.

Explicit Lorentz violation
is typically incompatible with Riemann geometry
\cite{akgrav,rbbid}.
Hence we consider the coefficient fields
as dynamical fields that spontaneously break Lorentz symmetry
via nonzero vacuum expectation values.
To provide a consistent treatment,
we also consider fluctuations about that vacuum value.
In Ref. \cite{lvgap}
the nature of the fluctuations for the coefficient fields of interest here
were established generally
in terms of the metric fluctuation $h_\mn$
and the vacuum values 
or \coef\ $\afb_\mu$ and $\cb_\mn$
for asymptotically flat spacetimes.
In keeping with standard practice in flat spacetime,
the \coef\ are taken to satisfy
$\prt_\al \afb_\mu =0$
and
$\prt_\al \cb_\mn =0$
in asymptotically inertial Cartesian coordinates.
The coefficients
$\afb_\mu$ and $\cb_\mn$
can then be identified with the \coef\
explored in Minkowski spacetime.
To define the vacuum values consistently with the Minkowski spacetime SME,
a constant $\al$ 
that characterizes couplings in the underlying theory 
was introduced that appears in front of the $\afb_\mu$ coefficients
in gravitational studies \cite{lvgap}.

As in the gravity sector,
the coefficient for Lorentz violation appearing here,
along with the masses and the gravitational constant,
are interpreted as effective quantities
by analogy with the EIH approach.
Note however
that other aspects of a full EIH approach including
tidal forces, multipole moments, and possible strong-field effects
lie beyond the present scope.

In the framework outlined above
it is found \cite{lvgap} that the Lorentz-violating contributions 
to the metric fluctuation
relevant for an analysis at third post-Newtonian order
take the form
\bea
h_{00} &\supset& \fr 2m [m\cb_{00} +  2 \al \afb_0]  U 
\nonumber\\
& & + \fr 2m [2 m\cb_{(j0)} + \al \afb_j] V^j
- \fr 2m \al \afb_j W^j,
\nonumber\\
h_{0j} &\supset& \fr 1m [\al \afb_j U
+ \al \afb_k U^{jk}]
\label{metric}
\eea
in harmonic coordinates.
Here the replacements \rf{composite}
can be used to obtain the metric fluctuation associated with a composite source
and 
$U$, $U^{jk}$, $V^j$, and $W^j$
are post-Newtonian potentials defined as
\bea
U &=& 
\G \int d^3 x^\prime 
\fr{\rh(\vec{x}^\prime, t)}
{|\vec{x}- \vec{x}^\prime|},
\\
U^{jk} &=& 
\G \int d^3 x^\prime 
\fr{\rh(\vec{x}^\prime, t) 
(\vec{x}- \vec{x}^\prime)^j 
(\vec{x}- \vec{x}^\prime)^k}
{|\vec{x}- \vec{x}^\prime|^3} ,
\nonumber\\
V^j &=& 
\G \int d^3 x^\prime 
\fr{\rh(\vec{x}^\prime, t) (v^{\rm S})^j (\vec{x}^\prime, t)}
{|\vec{x} - \vec{x}^\prime|},
\nonumber\\
\nonumber
W^j &=& 
\G \int d^3 x^\prime 
\fr{\rh(\vec{x}^\prime, t) 
(v^{\rm S})_k (\vec{x}^\prime, t) 
(\vec{x}- \vec{x}^\prime)^j 
(\vec{x}- \vec{x}^\prime)^k} 
{|\vec{x} - \vec{x}^\prime|^3}.
\label{potentials}
\eea
Here $\rh(\vec{x}^\prime, t)$ and $(v^{\rm S})_k$
are the density and velocity of the source respectively
in the relevant asymptotically inertial frame.
Further analysis \cite{lvgap} reveals that the relative acceleration 
of a pair of bodies,
here a pulsar and its companion,
 interacting gravitationally can be written
\bea
\fr{d^2 r^j}{dt^2} 
&=& 
- \fr{G}{r^3} 
\sum_{w} \Big[ 
M
+ 2 \nwr3 \al \afbw_0 r^j 
+ \nwr1  m^w \cbw_{00} r^j 
\nonumber \\
& & 
- 2 \et^{jk}  \nwr7  m^w \cbw_{(kl)} r^l 
+ 2 \nwr2 \al \afbw_k \et^{jk} v_{l} r^l 
\nonumber \\
& & 
- 2 \nwr2 \al \afbw_k v^{k} r^j 
- 2 \nwr6 m^w \cbw_{(0k)} v^k r^j 
\nonumber \\
& & 
+ 2 \et^{jk} ( \nwr6 - 2 \nwr8) m^w \cbw_{(0k)} v_l r^l  
\nonumber \\
& & 
- 2 \nwr8 m^w \cbw_{(0k)} v^j r^k \Big],
\label{ppacc}
\eea
to third post-Newtonian order
and leading order in the \coef.
Work here in the linearized limit
implies that indices may be raised and lowered with the Minkowski metric $\et_\mn$.
The relative position vector $\vec r = \vec r_1 - \vec r_2$,
where subscripts 1 and 2 denote the pulsar and companion respectively,
and $\vec v = \vec v_1 - \vec v_2$ is its coordinate-time derivative.
The system mass is $M = m_1 + m_2$,
$G$ is Newton's constant,
and the material-dependent factors $n_i^w$ are defined as follows:
\bea
\nwr1 &=&  
N^w_1 + N^w_2, 
\nonumber \\
\nwr2 &=& 
N^w_1 - N^w_2,
\nonumber \\
\nwr3 &=& 
M \left( \fr{N^w_1}{m_1} + \fr{N^w_2}{m_2} \right), 
\nonumber \\
\nwr4 &=& 
M \left( \fr{N^w_2}{m_2} - \fr{N^w_1}{m_1} \right),
\nonumber \\
\nwr5 &=& 
\fr{1}{M}(m_1 N^w_2 + m_2 N^w_1),
\nonumber \\
\nwr6 &=& 
\fr{1}{M} \left( m_1 N^w_2 - m_2 N^w_1 \right),
\nonumber \\
\nwr7 &=&  
\fr{m_2}{m_1} N^w_1 + \fr{m_1}{m_2} N^w_2,
\nonumber \\
\nwr8 &=& 
\fr{1}{M} \left( \fr{m^2_2}{m_1} N^w_1 
- \fr{m^2_1}{m_2} N^w_2 \right).
\label{compdepfact}
\eea

In many studies of modified gravity in pulsar systems,
the origin of the coordinates $(t,x^j)$ 
are taken to coincide with the conventional Newtonian center of mass
of the system, which is sufficient in many cases.
For example,
in the parameterized post-Newtonian formalism,
deviations that arise at post-Newtonian order 2 and beyond
are neglected in a typical analysis \cite{will}.
In the presence of anisotropic Lorentz violation,
the form of the conserved momentum and hence the form of the center of mass
is modified
in such a way that corrections to the center of mass typically arise at
lower post-Newtonian order.
Here we take the origin of the coordinates $(t,x^j)$ 
to coincide with the modified center of mass.
Consequences of the  modified center of mass
play a role in the analysis of Sec.\ \ref{tf}
and the details are developed there.
Relations between the frame $(t,x^j)$
and the Sun-centered frame used for reporting SME sensitivities \cite{data}
are provided in Sec.\ V E 5 of Ref.\ \cite{lvpn}.

\subsection{Keplerian Characterization}

\begin{figure}
\centerline{\psfig{figure=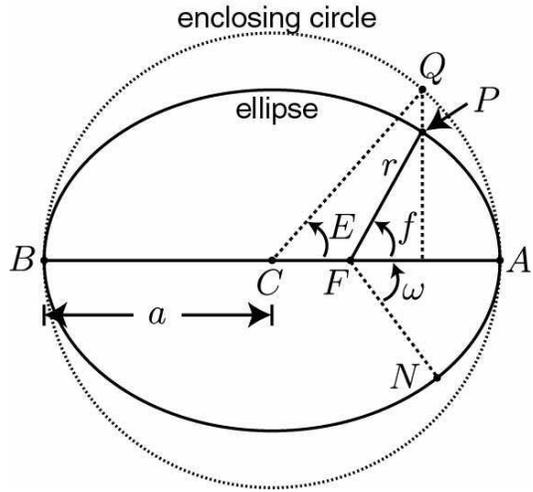,width=0.8\hsize}}
\caption{\label{f1} 
Variables describing the elliptical orbit.
Figure reproduced from Ref.\ \cite{lvpn}.}
\end{figure}

The standard keplerian characterization of an elliptical orbit
is used in the orbital analysis of binary-pulsar systems to follow.
The conventions are those of Ref.\ \cite{lvpn}.
Here we present the essential definitions
of the parameters describing the orbit,
and set up a relevant coordinate system for the analysis.

Some quantities helpful in developing the definitions to follow
are shown in Fig.\ \ref{f1}.
The focus of the ellipse is labeled F.
The point P located a coordinate distance $r$ from the focus
indicates the location of the pulsar at coordinate time $t$.
The semimajor axis of length $a$
lies between points B and C
and between points A and C.
The angle AFP is the true anomaly denoted $f$.
The eccentric anomaly $E$
is defined by the angle ACQ,
where Q is located at the intersection of the enclosing circle
and a line perpendicular to the major axis passing through P.
The angle $\om$ is from the line of ascending nodes FN
to the major axis.

The analysis to follow makes use of the general solution
of the elliptic two-body problem.
The notation $n$ is used for the orbital frequency
and the period is defined as usual $P_b = 2\pi/n$.
The solution is generally characterized
by six constants known as orbital elements:
the semimajor axis $a$,
the eccentricity $e$,
the inclination of the orbit $i$,
the longitude of the ascending node $\Om$,
the angle $\om$ introduced above,
and the mean anomaly $l_0$ at the epoch $t=t_0$
related to the eccentric anomaly $E$ via
\beq
E-e \sin E = l_0 + n (t-t_0).
\label{mean anomaly}
\eeq
Further details can be found in 
Refs.\ \cite{lvpn,osc}.

\begin{figure}
\centerline{\psfig{figure=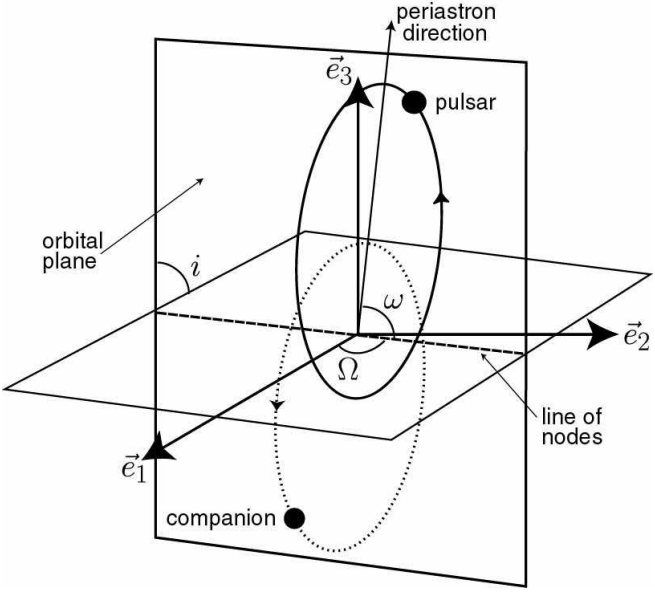,width=\hsize}}
\caption{\label{bp} 
Elliptical orbits in the post-Newtonian frame of the binary pulsar.
Figure reproduced from Ref.\ \cite{lvpn}.}
\end{figure}

The orbital elements $i$, $\Om$, $\om$ 
specify the orientation of the orbit in the reference coordinate system
as shown in Fig.\ \ref{bp}.
It is also convenient to define a set of unit vectors
that specify the orientation of the orbit.
A typical choice defines
$\vec k$ normal to the plane of the orbit,
$\vec P$ pointing from the focus to the periastron,
and $\vec Q = \vec k \times \vec P$.
The relation between these vectors
and the basis of the reference coordinate system ${\bf e}_j$ is then
\bea
\vec P &=& (\cos \Om \cos \om - \cos i \sin \Om \sin \om) \vec e_1
\nonumber\\
&&
+ (\sin \Om \cos \om + \cos i \cos \Om \sin \om) \vec e_2
\nonumber\\
&&
+ \sin i \sin \om \vec e_3,
\nonumber\\
\vec Q &=& -(\cos \Om \sin \om + \cos i \sin \Om \cos \om) \vec e_1
\nonumber\\
&&
+ (\cos i \cos \Om \cos \om -\sin \Om \sin \om) \vec e_2
\nonumber\\
&&
+ \sin i \cos \om \vec e_3,
\nonumber\\
\vec k &=& \sin i \sin \Om \vec e_1
- \sin i \cos \Om \vec e_2
\nonumber\\
&&
+ \cos i \vec e_3.
\label{k}
\eea
With these definitions,
the unperturbed orbit can be written
\bea
\vec r &=& \fr {a(1-e^2)}{1+e \cos f} (\vec P \cos f + \vec Q \sin f).
\label{r}
\eea

\section{Secular Changes}
\label{sc}

Nonzero coefficients for Lorentz violation 
$(\bar{a}_{\rm{eff}})_{\mu}$ and $\bar{c}_{\mu v}$
affect the pulsar and the companion's relative motion 
and therefore modify the original Newtonian elliptical orbit. 
This section characterizes these effects
by obtaining the secular changes in the six orbital elements 
$a,e, l_0, i, \Omega, \omega$ generated by the \coef.

To proceed, 
we adopted the method of osculating elements \cite{lvpn,osc}, 
which is a standard method 
that treats the relative motion of the pulsar and its companion 
as instantaneously part of an unperturbed elliptical orbit.
Here this unperturbed orbit solves the equation
\beq
\fr{d^2 r^j}{dt^2} = 
- \fr{G M \la}{r^3} r^j,
\label{unperturbed}
\eeq
where
\bea
\lambda=1+\fr{1}{M}\sum_wn_1^wm^w \cbw_{00}+ \fr{2}{M}\sum_wn_3^w\left(\bar{a}^w_{\rm eff}\right )_{0}.
\eea
The relation between the orbital frequency $n$ and the semimajor axis $a$
then takes the form
\beq
n^2 a^3 = G M \la.
\eeq
Note that the combination of coefficients $\la$ appears
as an effective contribution to $G M$ and is unobservable
via the analysis of a single system.

Perturbations to the orbit
are then characterized by a gradual time-evolution of the ellipse,
with the six orbital elements
now being functions of time.
The perturbing acceleration can now be 
defined as the difference between the full modified acceleration \rf{ppacc} 
and the central acceleration \rf{unperturbed}.
The result takes the form
\bea
\nonumber
\mathrm{a}'^{j}= \fr{d^2r^j}{dt^2} - \fr{d^2r^j_0}{dt^2} = 
\fr{G}{r^2} \Big( A^j_{\phantom{j}l} \hat{r}^l
+B^j v_l \hat{r}^l +C^k v_k \hat{r}^j\\
+\fr{1}{2}(B^k+C^k)\hat{r}_k v^j\Big),
\label{pertacc}
\eea
where we classify the combinations of $\afb_\mu$ and $\cb_\mn$ into 3 kinds, 
namely $A^j_{\phantom{j}l}$, $B^j$, and $C^j$, 
according to their relations with $\vec{v}$ and $\vec{r}$ in \rf{ppacc}. 
Specifically, 
\bea
A^j_{\phantom{j}l}&=& \sum_w 2\eta^{jk}n_7^wm^w \cbw_{\left(kl\right )}\nonumber \\
B^j &=& - \sum_w 2\eta^{jk} \Big(n_2^w\alpha\left(\bar{a}_{\rm eff}^w\right )_k\nonumber \\
& &
+ \left(n_6^w-2n_8^w\right )m^w \cbw_{\left(0k\right )}\Big)\nonumber \\
C^j &=& \sum_w  2\eta^{jk} \Big( n_2^w\alpha\left(\bar{a}_{\rm eff}^w\right )_k
+ n_6^wm^w \cbw_{\left(0k\right ) }\Big).
\eea

Following the method of osculating elements \cite{osc}
then requires the following steps.
The relative position $\vec r$ from Eq.\ \rf{r}
along with its coordinate-time derivative $\vec v$
are inserted into the perturbing acceleration \rf{pertacc}.
The relevant projections of the perturbing acceleration
are then used to extract the variations in the orbital elements.
The variations are averaged over an orbit
via integration over the true anomaly $f$. 
For the secular variation in the semimajor axis
and the eccentricity we find
\bea
\Big \langle\fr{da}{dt} \Big \rangle &=& 0 \nonumber \\
\Big \langle\fr{de}{dt}\Big \rangle &=&  \fr{n}{M}\left(1-e^2\right )^{1/2}\Big(\fr{e^2-2\varepsilon}{e^3}A_{PQ}+\fr{na\varepsilon}{e^2}B_P \Big), \phantom{aa}
\label{evole}
\eea
where $\ep=1- (1-e^2)^{1/2}$ is the eccentricity function. 
The secular changes in the orbital elements related to the orientation of the ellipse 
are given by
\bea
\Big \langle\fr{di}{dt}\Big \rangle = \hspace{-.1in} && \fr{n}{M\left(1-e^2\right )^{1/2}}\Big(\fr{\varepsilon}{e^2}A_{kP}\cos\omega-\fr{e^2-\varepsilon}{e^2}A_{kQ}\sin\omega \nonumber \\
& &  
-\fr{n\varepsilon a}{e}B_{k}\sin \omega\Big ) \nonumber \\
\Big \langle\fr{d\Omega}{dt}\Big \rangle = \hspace{-.1in} && \fr{n \sin i}{M\left(1-e^2\right )^{1/2}}\Big(\fr{\varepsilon}{e^2}A_{kP}\sin\omega +\fr{e^2-\varepsilon}{e^2}A_{kQ}\cos\omega \nonumber \\
& &
+\fr{n\varepsilon a}{e}B_{k}\cos \omega\Big ) \nonumber \\
\Big \langle\fr{d\omega}{dt}\Big \rangle = \hspace{-.1in} && -\fr{n}{M\tan i\left(1-e^2\right )^{1/2}}\Big(\fr{\varepsilon}{e^2}A_{kP}\sin\omega+ \nonumber \\
& & 
\fr{e^2-\varepsilon}{e^2}A_{kQ}\cos\omega+\fr{n\varepsilon a}{e}B_{k}\cos \omega\Big ) \nonumber \\
& & 
+\fr{n}{M}\Big[\fr{e^2-2\varepsilon}{2e^4}\left(A_{QQ}-A_{PP}\right ) \nonumber \\
&& \phantom{+\fr{n}{M}\Big[} +\fr{na\left(e^2-\varepsilon\right )}{e^3\left(1-e^2\right )^{1/2}}B_Q\Big],
\space 
\eea
while secular evolution of the mean anomaly at the epoch
takes the form
\bea
\Big \langle\fr{dl_0}{dt}\Big \rangle &=& -\fr{n}{M}\Big[\fr{(e^2-2\varepsilon)\left(1-e^2\right )^{1/2}}{2e^4}\left(A_{QQ}-A_{PP}\right ) \nonumber \\
& & 
+\fr{na\left(e^2-\varepsilon\right )}{e^3}B_Q
+\fr{2\varepsilon}{e^2}A_{PP}+\fr{2(e^2-\varepsilon)}{e^2}A_{QQ}
\nonumber \\
& &
+\fr{3na\varepsilon}{e}\left(B_Q+C_Q\right)\Big].
\label{evolma}
\eea
Notice that these results are not limited to binary-pulsar systems 
but could be applied in any two-body celestial system.
The application of these results
to searches for Lorentz violation in binary pulsars
is considered in the following sections.

Before concluding our development of secular changes
in pulsar systems,
we note that secular changes in the spin of solitary pulsars
also arises in the presence of matter-sector \coef.
The effect can be attributed to the violations
of angular-momentum conservation that accompany Lorentz violation.
Following the derivation in Ref.\ \cite{Spin},
we find
\beq
\Omega_k^{prec}=-\fr{2\pi}{m P} \sum_w N^w m^w \cbw_{jk} \hat{S}^j
\eeq
for the spin-precession frequency of a solitary pulsar
at second post-Newtonian order.
Here $P$ is the spin period,
$m$ is the mass of the spinning body,
and $\hat{S}$ is the unit vector pointing along the spin direction. 
Since spin precession can also be extracted from the pulse data, 
it is then also possible to measure the matter-sector coefficients for Lorentz violation 
in the context of solitary pulsars
as well as other spinning bodies.
This idea has already been used in the context of searches
for changes in the magnitude of the angular velocity of pulsars
to place constraints on combinations of $\cbn_{jk}$ \cite{brettpuls}.

\section{Timing formula}
\label{tf}

The secular changes in the orbit of the pulsar system
are observable through the rate at which pulses are received at Earth.
To complete the prediction,
the secular changes in the nature of the orbit
should be combined with other curved-spacetime effects
on the pulse rate.
Here we consider such effects in developing a timing formula
to be used in conjunction with the orbital elements above.

We begin by considering the trajectory of a photon
traveling from the pulsar to Earth.
The trajectory is determined by the null condition
$ds^2=0$ where the relevant metric is obtained from \rf{metric}.
The same simplifications used in the gravity sector \cite{lvpn}
hold here.
Time-delay effects associated with the pulsar itself
can be disregarded in this analysis since they result in a constant shift.
Time-delay effects from the companion,
which could be obtained via the general analysis of time delay in the matter sector \cite{lvgap},
are negligibly small in the present context.
Effects of solar-system bodies and the motion of the detector
are omitted for simplicity.
With this,
the photon trajectory reduces to the simple relation
\beq
t_{\rm arr} - t_{\rm em} =
|\vec r_E - \vec r_1|,
\label{tarr}
\eeq
where $\vec r_E$ is the location of Earth,
$t_{\rm em}$ is the coordinate time of emission by the pulsar,
and $t_{\rm arr}$ is the coordinate time of arrival at Earth.

Contact must then be made with the proper time of emission.
This can be done through consideration of the relation between the proper-time interval
measured by an ideal clock at the surface of the pulsar and the coordinate-time interval
obtained from the metric \rf{metric}
\bea
d\tau = dt[1-kU-\half \vec{v}_1^2].
\label{dtau}
\eea
Here $\vec{v}_1$ is the velocity of the pulsar.
The potential in \rf{dtau} includes contributions from both the pulsar 
and its companion evaluated along the trajectory of the clock $U=\fr{Gm_2}{r}+U_1 $.
The contribution $U_1$ from the pulsar 
generates an unobservable constant shift and will be neglected in what follows.
As a result,
$k$ can be taken as the isotropic combination of SME coefficients 
\bea
k=1+\fr{1}{m_2} \sum_w N_2^wm^w \cbw_{00}+\fr{2}{m_2}\sum_w N_2^w\left(\bar{a}_{\rm eff}^w\right )_{0}.
\eea
The velocity of the pulsar $\vec{v}_1$ can be obtained from the relative velocity  
$\vec{v}$ through the modified version of conservation of momentum 
at second post-Newtonian order
along with our coordinate choices.
The total momentum of the system takes the form
\bea
(p_{\rm Tot})_j &=& 
m_1(\delta_{jk} + \cb^1_{00} \de_{jk} + 2\bar{c}^1_{jk})(v_1 )_k
\nonumber\\
& & + m_2(\delta_{jk} + \cb^2_{00} \de_{jk} +2\bar{c}^2_{jk})(v_2)_k
\nonumber\\
& & + (p_a)_j
+ (p_c)_j 
+ (p_s)_j + \ldots,
\label{ptotal}
\eea
where a superscript $1$ indicates the coefficients for the pulsar 
and $2$ indicates the coefficients for its companion,
which can be expanded in terms of the species content of the bodies
via Eq.\ \rf{composite}. 
The additional terms 
\bea
\nonumber
(p_a)_j &=& \fr{2 G \al}{r} \left( m_1 \afbx2_j + m_2 \afbx1_j \right)\\
(p_c)_j &=& \fr{m_1 m_2^2}{M^2} \left( \cbx1_{0j} v^2 + 2 \cbx1_{0k} v_k v_j \right)
+ \fr{ 2 G m_1 m_2}{r} \cbx1_{0j}
\nonumber \\
& & 
+ (1 \leftrightarrow 2)
\eea
denote contributions at post-Newtonian order three
that are relevant for our subsequent analysis.
We note in passing that a similar order three contribution exists
associated with gravity-sector coefficient $\sb^\mn$:
\beq
(p_s)_j = \fr{G m_1 m_2}{r} \left( 3 \sb_{0j} + \sb_{0k} \hat r_k \hat r_j \right).
\eeq
Our choice of frame is fixed by the condition $(p_{\rm Tot})_j = 0$.

We then integrate \rf{dtau} to obtain the desired relation
between the coordinate time of emission and the proper time of emission.
The integration is facilitated by a change variables from coordinate time 
to the eccentric anomaly $E$, 
which is related to the true anomaly $f$ by 
\bea
\tan \fr{f}{2}=\left(\fr{1+e}{1-e}\right)^{1/2}\tan \fr{E}{2}.
\label{ftoE}
\eea
Replacing $r$ in \rf{dtau} using \rf{r} along with \rf{ftoE}, 
we can then perform integration over $E$,
which yields,
up to constants,
\bea
\tau = && \hskip-8pt t 
- \fr{kGm_2P_b}{2\pi a}E
-\fr{\lambda Gm_2^2P_b}{2M\pi a}E
- \sum_w \fr{Gm_1 m_2 m^w n_4^w P_b}{M^2 \pi a} \nonumber \\
&& \times \Big[\fr{m_2 \cbw_{00}}{M} E
- 2 \fr{\cbw_{PP}-\left(1-e^2\right ) \cbw_{QQ}}{e^2}E \nonumber \\
&& +\left( \cbw_{PP}- \cbw_{QQ}\right )\mathcal{K}_1\left(E\right )
- \cbw_{PQ} \mathcal{K}_2\left(E\right )\Big].
\label{tautem}
\eea
Subscripts $P$, $Q$  and $k$ on the \coef\
indicate projections along the directions $\vec{P}$ , $\vec{Q}$ and $\vec{k}$.
For example, $\cbw_P=-\cbw_{0j}P_{j}$. 
The functions $\mathcal{K}_1$ and $\mathcal{K}_2$ appearing in (14) are given by
\bea
\mathcal{K}_1\left(E\right ) &=& \fr{2\left(1-e^2\right )^{1/2}}{e^2}\arctan\left[\left(\fr{1+e}{1-e}\right )^{1/2}\tan \frac{1}{2}E\right] \nonumber \\
\mathcal{K}_2\left(E\right ) &=& \fr{2\left(1-e^2\right )^{1/2}}{e^2}\Big[e\cos E+\ln \left(1-e\cos E\right )\Big].
\eea
The constant in Eq.\ \rf{mean anomaly}
is fixed via the choice $E-e \sin E = n(t_{\rm arr} -r_E)$.

The desired relation between the proper time of emission
and the time of arrival is then obtained by combining Eq.\ \rf{tautem}  with Eq.\ \rf{tarr}
yielding
\bea
\ta &=& t_{\rm arr} - \cA (\cos E-e) 
- \cB \sin E 
\nonumber\\
&&
- \fr{2 \pi a a_1 e}{P_b} \left( k + \frac{m_2}{M} \la \right) \sin E
\nonumber\\
&&
- \fr{2\pi}{P_b}
\fr{(\cA \sin E- \cB \cos E)}
{(1-e \cos E)}
[\cA(\cos E-e) + \cB \sin E] 
\nonumber\\
&&
- \sum_w \fr{Gm_1 m_2 m^w n_4^w P_b}{M^2 \pi a} 
\Big[\fr{e m_2 \cbw_{00}}{M} \sin E \nonumber \\
&& - 2 \fr{\cbw_{PP}-\left(1-e^2\right ) \cbw_{QQ}}{e} \sin E 
 +\left( \cbw_{PP}- \cbw_{QQ}\right )\cK_1(E)\nonumber \\
&& - \cbw_{PQ} \cK_2(E)\Big] + \ta_{{\rm CM} a} + \ta_{{\rm CM} c}.
\label{tautarr}
\eea
Here $\cA$ and $\cB$ are the combinations
\bea
\cA &=& a_1 \sin i \sin \om,
\nonumber\\
\cB &=& \sqrt{(1-e^2)} ~a_1 \sin i \cos \om,
\eea
and $a_1=a m_2/M$.
Note that with the exception of Eq.\ \rf{mdef},
the \coef\ $a_\mu$ appears in the combination
$\af_\mu$ throughout this work.
Hence conflict between these two standard uses
of a lower index on an object $a$ is avoided.
In obtaining Eq.\ \rf{tautarr},
our choice of origin as the modified center of mass
is used to fix $\vec r_1$. 
Explicitly our choice corresponds to the requirement
\beq
\int (p_{\rm Tot})_j dt = 0.
\eeq
The contributions $\ta_{{\rm CM} a}$ and $\ta_{{\rm CM} c}$
are a result of the Lorentz violating contributions to the momentum
associated with the $\afb_\mu$ and $\cb_\mn$ coefficients respectively.
The contribution associated with $\afb_\mu$ takes the form
\bea
\ta_{{\rm CM} a} &=& 
-\fr{4 \pi a e}{P_b m_2} \sum_w n^w_5 \al \sin E \Big[ \afw_P \cA
\\ && \hskip20pt
+ \afw_Q \fr{\cB}{\sqrt{1-e^2}} 
+ \afw_k a_1 \cos \om \Big],
\nonumber 
\eea
while $\cb_\mn$ contributes
\bea
\ta_{{\rm CM} c} \hskip-8pt && = - \sum_w \fr{2 m_1 m^w n^w_4}{M^2} 
[ \Xi^w (\cos E -e) + \Si^w \sin E ]
\nonumber \\ & & 
- \sum_w \fr{4 \pi m_1 m^w n^w_4}{P_b M^2} 
\nonumber \\ & & 
\times \Big[ \fr{\cA (\cos E - e) + \cB \sin E}{1 - e \cos E} 
\left( \Xi^w \sin E - \Si^w \cos E \right)
\nonumber \\ & & \hskip5pt
+ \fr{\cA \sin E - \cB \cos E}{1 - e \cos E} 
\left( \Xi^w (\cos E - e) + \Si^w \sin E \right)\Big]
\nonumber \\ & & 
+ \sum_w \fr{4 \pi a m_1 m^w}{P_b M^2}
\Big[(n^w_3 + 2 n^w_7) \Up^w e \sin E 
\nonumber \\ & & \hskip5pt
+ \half n^w_7 \Big( \cbw_Q \cA + \cbw_p \fr{\cB}{\sqrt{1-e^2}} \Big) \cK_2(E)
\nonumber \\ & & \hskip5pt
+ n^w_7 \Big( \cbw_P \cA - \cbw_Q \fr{\cB}{\sqrt{1-e^2}} \Big) (\cK_1(E) - \frac{2}{e} \sin E) \Big]. 
\eea
Here $\Xi^w$ and $\Si^w$ are defined as
\bea
\nonumber
\Xi^w &=& (\half \cbw_{00} + \cbw_{PP}) \cA +  \cbw_{PQ} \fr{\cB}{\sqrt{1-e^2}} +  \cbw_{Pk} a_1 \cos i \\
\nonumber
\Si^w &=& \cbw_{QP} \cA \sqrt{1-e^2} + (\half \cbw_{00} + \cbw_{QQ}) \cB + \cbw_{Qk}  a_1 \cos i\\
\Up^w &=& \cbw_P \cA +  \cbw_Q \fr{\cB}{\sqrt{1-e^2}} +  \cbw_k a_1 \cos i.
\eea
We note that the result of including this effect in the gravity sector
would generate the additional contribution
\bea
\ta_{{\rm CM} s} &=& - \fr{2 \pi a m_1}{P_b M}  \Big[
3 e \sb_k a_1 \cos i \sin E
\nonumber \\
& & + \sb_P \Big (\cA \fr{(1+3e^2) \sin E}{e} + \cA \cF_1(E) + \cB \cF_2(E)\Big) 
\nonumber \\
& & - \fr{\sb_Q}{\sqrt{1-e^2}} \Big (\cB \fr{(1-4e^2) \sin E}{e} - \cB \cF_1(E) 
\nonumber \\
& &
\phantom{- \fr{\sb_Q}{\sqrt{1-e^2}} \Big (}
- (1-e^2) \cA \cF_2(E) \Big) \Big]
\eea
in the arrival time relation for the gravity sector,
Eq.\ (180) of Ref.\ \cite{lvpn}.
The functions $\cF_1(E)$ and $\cF_2(E)$ are provided
as Eq.\ (178) of Ref.\ \cite{lvpn}.

The results of this section,
used in conjunction with the evolution of the orbital elements
could then be used to search for matter-sector Lorentz violation
in pulsar timing data.
A direct search could be done by inserting Eq.\ \rf{tautarr}
into a model for the emission of pulses as a function
of the proper time at the surface of the pulsar $\ta$.
This would provide an equation for the received pulses
as a function of arrival time
that could be fit to the observational data.
Limits can also be placed
through the statistical analysis of 
existing fits to a number of pulsar systems
\cite{shao},
such as fits performed in the parametrized post-Keplerian formalism
\cite{ppk}.

Crude estimates of the sensitivities that might be obtained
via pulsar systems can be seen by comparing the results
of Shao's recent analysis of the pure-gravity sector \cite{shao}
with existing sensitivities to $\afb_\mu$ and $\cb_\mn$ \cite{data}.
For example,
Shao finds sensitivity to $\sb^{TJ}$ at the $10^{-8}$ level.
Based on the similarity in the equations,
sensitivity to $\afb_J$ at a similar level might be expected,
which would amount to an improvement of about 3 orders of magnitude
over existing constraints.
It should also be emphasized that in the matter sector,
different systems having different compositions
offer different sensitivities.
For example,
a neutron star-white dwarf system
is sensitive to a different linear combination
of coefficients than a double neutron star system.
If the existence of more exotic stars is confirmed,
such as strange stars,
these systems may offer unique sensitivities to the associated coefficients.
The prospects noted here
are likely to be further enhanced by anticipated observational improvements
in the coming years.

\section{Summary}
\label{summary}

In this work we have derived the effects
of matter-sector Lorentz violation on both the motion
of binary-pulsar systems
and the pulses traveling from them
via a post-Newtonian treatment.
The Lorentz-violating corrections to the relative motion
of the pulsar and companion are developed in Sec.\ \ref{sc}
via the method of osculating elements.
Equations \rf{evole}-\rf{evolma}
provided the secular changes to the orbital elements
resulting from this analysis.
These results can then be used in conjunction
with the modifications to the timing formula
developed in Sec.\ \ref{tf},
where the central result
is the relation between the proper time of pulse emission
and the time of arrival \rf{tautarr}.
In obtaining this result,
we consider
the effects of Lorentz-violating corrections
to the conserved momentum \rf{ptotal}
in pulsar systems.

Analysis of existing pulsar-timing data
using the results presented here
will provide the first consideration
of Lorentz violation in matter-gravity couplings
in strong-gravity systems,
and
will yield significant sensitivity improvements
over existing solar system and laboratory sensitivities.
Matter-sector Lorentz violation
is in general accompanied by particle-species dependent effects,
making measurements in systems involving different types of bodies
of significant interest.
The use of existing data can provided about 3 orders of magnitude improvement
over existing measurements from other systems.
Going forward,
prospects for additional improvement
via observational advances are excellent.

\section*{Acknowledgments}
\label{Acknowledgments}

This work was supported in part 
by Carleton College Towsley Funds.
We acknowledge useful discussion
with L.\ Shao.

\end{document}